# Opto-Valleytronic Spin Injection in Monolayer MoS$_2$/Few-Layer Graphene Hybrid Spin Valves

Yunqiu (Kelly) Luo,[1#] Jinsong Xu,[1#] Tiancong Zhu,[1] Guanzhong Wu,[1] Elizabeth J. McCormick,[1] Wenbo Zhan,[1] Mahesh R. Neupane,[2] Roland K. Kawakami[1*]

[1]*Department of Physics, The Ohio State University, Columbus, OH 43210, USA*

[2]*Sensors and Electron Devices Directorate, U.S. Army Research Laboratory, Adelphi, Maryland 20783, USA*

[#]equal contributions

*Corresponding Author

   e-mail:    kawakami.15@osu.edu

   Address:   191 W. Woodruff Ave.
                     Department of Physics
                     The Ohio State University
                     Columbus, OH  43210

   Phone:    (614) 292-2515

   Fax:      (614) 292-7557

**ABSTRACT**

Two dimensional (2D) materials provide a unique platform for spintronics and valleytronics due to the ability to combine vastly different functionalities into one vertically-stacked heterostructure, where the strengths of each of the constituent materials can compensate for the weaknesses of the others. Graphene has been demonstrated to be an exceptional material for spin transport at room temperature, however it lacks a coupling of the spin and optical degrees of freedom. In contrast, spin/valley polarization can be efficiently generated in monolayer transition metal dichalcogenides (TMD) such as $MoS_2$ via absorption of circularly-polarized photons, but lateral spin or valley transport has not been realized at room temperature. In this letter, we fabricate monolayer $MoS_2$/few-layer graphene hybrid spin valves and demonstrate, for the first time, the opto-valleytronic spin injection across a TMD/graphene interface. We observe that the magnitude and direction of spin polarization is controlled by both helicity and photon energy. In addition, Hanle spin precession measurements confirm optical spin injection, spin transport, and electrical detection up to room temperature. Finally, analysis by a one-dimensional drift-diffusion model quantifies the optically injected spin current and the spin transport parameters. Our results demonstrate a 2D spintronic/valleytronic system that achieves optical spin injection and lateral spin transport at room temperature in a single device, which paves the way for multifunctional 2D spintronic devices for memory and logic applications.

Spintronics and valleytronics, novel fields with large potential impacts in both fundamental science and technology, utilize the electron's spin and valley degrees of freedom, in addition to charge, for information storage and logic operations. In the past decade, experimental studies have established single-layer and multilayer graphene as among the most promising materials for spintronics due to their high electronic mobility combined with low intrinsic spin-orbit coupling. Graphene exhibits room temperature spin diffusion length of up to tens of microns, substantially longer than conventional metals or semiconductors (<1 micron)[1-4]. However, graphene's lack of spin-dependent optical selection rules has made opto-spintronic functionality impossible, a substantial limitation for graphene.

Fortunately, monolayer $MoS_2$ and related semiconducting transition metal dichalcogenides (TMDs) exhibit favorable characteristics for nanoscale opto-valleytronic and opto-spintronic applications[5-7]. TMDs have strong spin-orbit coupling due to the heavy metal atom and lack inversion symmetry in monolayer form, the combination of which allows complete simultaneous valley and spin polarization through absorption of circularly polarized light[8-14]. This originates from the valley-dependent optical selection rules of monolayer $MoS_2$, where absorption of circularly polarized σ+ (σ-) photons excites electrons only in the K (K′) valley. Because this valley selection rule derives from the symmetries of the lattice, it is a general rule that also applies to systems with low SO coupling such as monolayer hBN and gapped graphene[9, 10] where the valley-dependent optical transition is independent of spin. In monolayer $MoS_2$, however, the spin selection is induced by the strong SO coupling. In the K (K′) valley, the valence band has a large spin-orbit splitting with a spin up (down) state at the valence band maximum and spin down (up) state lower in energy, with SO splitting of ~150 meV[8, 15]. Therefore, the spin and valley degrees of freedom are strongly coupled, and the valley optical selection rule can be used to generate spin-polarized photoexcitation.

The true strength of graphene and TMDs for spin- and valleytronics lies in the combination of the two materials, where the strengths of each material can compensate for the weaknesses of the other. It has already been demonstrated that manipulation of spin currents in graphene is possible through proximity to TMDs via spin absorption[16, 17], as well as proximity to magnetic insulators through exchange fields[18, 19]. Additionally, Fabian and co-workers proposed that absorption of circularly-polarized photons in monolayer $MoS_2$ will create valley/spin polarized excitations that can generate spin injection into an adjacent graphene layer[20]. This would provide a route toward opto-spintronic functionality in graphene by creating a vertical heterostructure with monolayer $MoS_2$.

In this Letter, we experimentally demonstrate spin injection from monolayer $MoS_2$ to few-layer graphene following optical valley/spin excitation in $MoS_2$ with circularly polarized light. We detect spins in graphene through voltage signals on a ferromagnetic (FM) electrode in a non-local measurement geometry. Notably, the spins in graphene precess in an external magnetic field and we obtain antisymmetric Hanle spin precession curves, which prove that the measured voltage signals originate from optical spin injection and spin transport. In addition, we find that tuning the photon energy adjusts the magnitude and direction of the injected spin polarization, which is a direct consequence of the large spin splitting in the valence band of $MoS_2$. Low temperature measurements (10 K) reveal a double peak structure in the spin signal spectrum near the A exciton resonance, while measurements at elevated temperatures find that the opto-valleytronic spin injection into graphene persists up to room temperature. Lastly, we quantify the injected spin current using a one-dimensional spin transport model based on the Bloch equations. Our results demonstrate unprecedented spintronic/valleytronic functionality of a TMD/graphene device by integrating opto-valleytronic spin injection, lateral spin transport and electrical spin detection in a single van der Waals heterostructure.

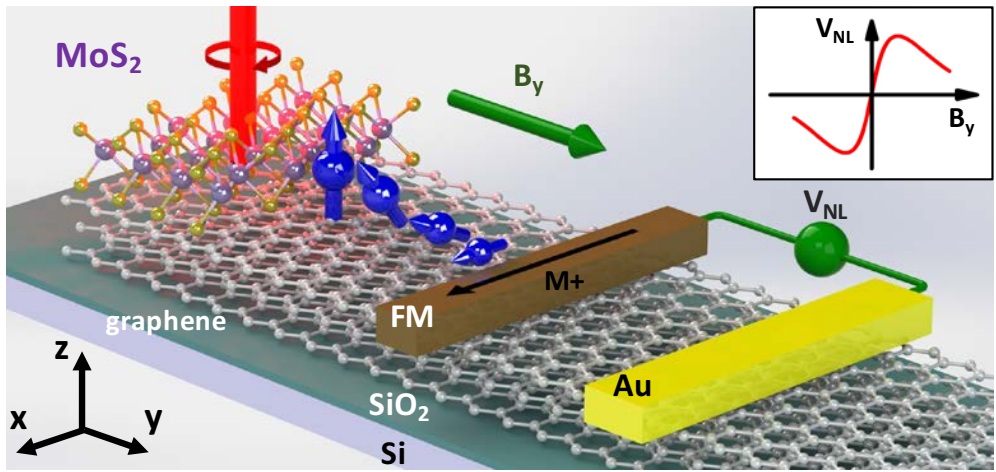

**Figure 1.** Illustration of optical spin injection, lateral spin transport and electrical spin detection in a monolayer $MoS_2$/few-layer graphene hybrid spin valve structure. Inset: expected signal $V_{NL}$ as a function of applied magnetic field $B_y$.

As illustrated in Figure 1, the concept of the experiment is to optically excite spin/valley polarization in $MoS_2$ in order to inject spin polarization into the underlying graphene, where it diffuses and precesses in an external magnetic field, and is finally detected electrically by a FM electrode. We begin with the absorption of circularly polarized photons in monolayer $MoS_2$ to produce spin/valley-polarized carriers oriented out-of-plane (along +z), which subsequently transfer into the adjacent few-layer graphene. The spins (blue arrows) then diffuse within the few-layer graphene toward a ferromagnetic (FM) spin detector with in-plane magnetization. To detect the spin transport, a magnetic field $B_y$ is applied to induce spin precession. This generates a non-zero component of spin-polarization ($S_x$) along the FM detector's magnetization, which produces a detector voltage ($V_{NL}$) that is proportional to $S_x$. By measuring $V_{NL}$ as a function of $B_y$, the combined processes of optical spin injection, lateral spin transport, and electrical spin detection can be identified as an antisymmetric Hanle curve as shown schematically in the Figure 1 inset.

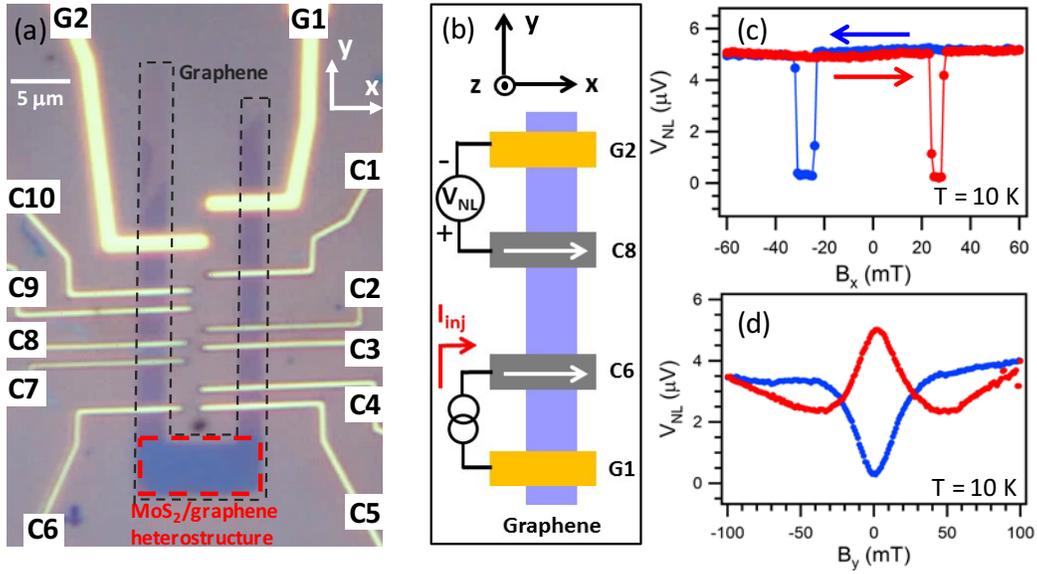

**Figure 2.** Electrical spin transport measurements in few-layer graphene. (a) Optical microscope image of a monolayer $MoS_2$/few-layer graphene hybrid spin valve. The black dashed line highlights the few-layer graphene region. The red dashed line highlights the monolayer $MoS_2$/graphene junction region. C1-10 are cobalt electrodes. G1 and G2 are gold electrodes. (b) Schematic of the non-local spin valve measurement. (c) Non-local spin valve measurement. The red (blue) curve is for the up (down) sweep of magnetic field $B_x$ parallel to the Co magnetization. (d) In-plane Hanle spin precession measurement. The red (blue) curve is for the parallel (antiparallel) alignment of the Co magnetizations.

To realize this experimentally, we fabricate a monolayer $MoS_2$/few-layer graphene hybrid spin valve, shown in Figure 2a. This device consists of n-type few-layer graphene (black dashed lines) contacted by monolayer $MoS_2$ (red dashed lines), Cr/Au electrodes (G1, G2) and Co electrodes with SrO tunnel barriers (C1-C10). Details of sample fabrication and material characterization are in the Supporting Information (SI), Sec. 1. Before attempting optical spin injection, we first establish the proper electrical spin injection, transport and detection processes in few-layer graphene using the non-local magnetotransport geometry at 10 K, as shown in Figure 2b. The current $I_{inj}$ (= 1 µA) injects spin polarized electrons into graphene at injector electrode C6. The spins subsequently diffuse in graphene towards the spin detector C8, where it is measured as a voltage signal $V_{NL}$ across electrodes C8 and G2 (nonmagnetic reference electrode). Figure 2c shows $V_{NL}$ as a function of magnetic field applied parallel to the Co electrodes ($B_x$), resulting in hysteretic jumps as the Co magnetizations switch between parallel (high $V_{NL}$) and antiparallel (low $V_{NL}$) configurations. The presence of these jumps in $V_{NL}$ indicates spin transport through graphene. To extract the spin transport parameters of the few-layer graphene, we perform in-plane Hanle spin precession measurements by applying a magnetic field perpendicular to the electrode axis ($B_y$). The measured $V_{NL}$ for parallel (red circles in Fig. 2d) and antiparallel (blue circles in Fig. 2d) states are analyzed to yield a spin lifetime of $\tau_G$ = 308 ps, diffusion coefficient $D_G$ = 0.0301 m$^2$/s, and spin diffusion length of $\lambda_G = \sqrt{D_G \tau_G}$ = 3.04 μm (see SI, Sec. 2 for details of the spin transport measurement and analysis).

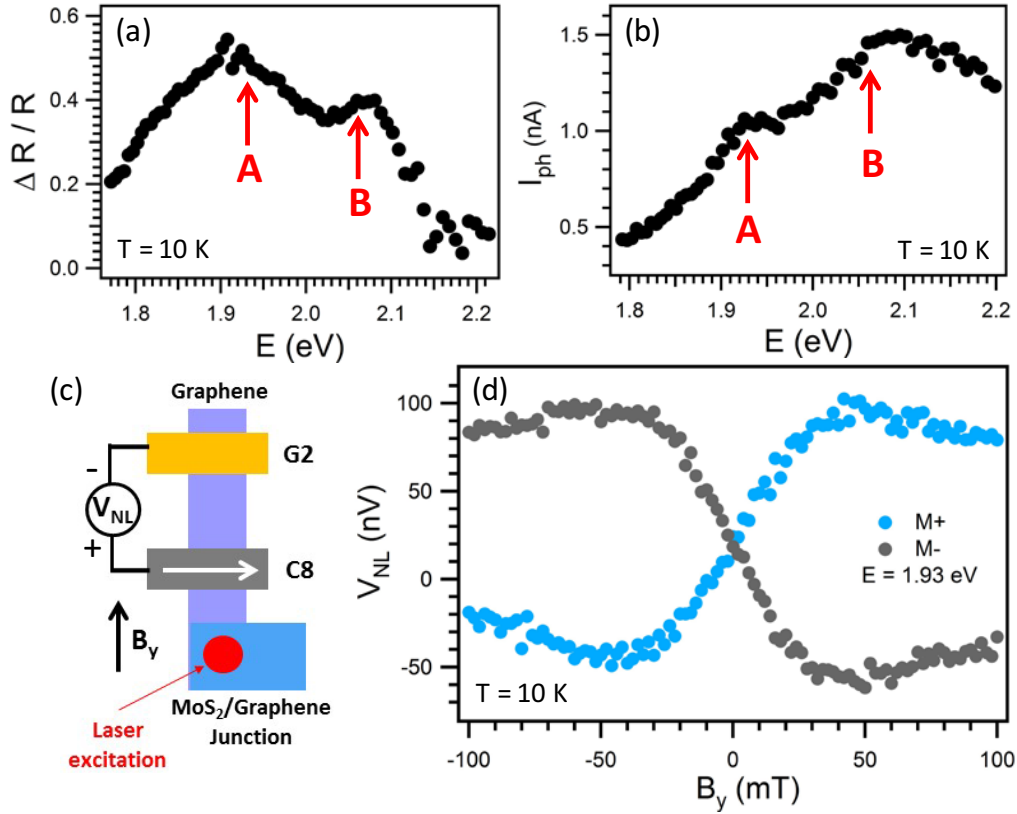

**Figure 3.** Electrical spin detection of the opto-valleytronic spin injection. (a) Reflection contrast spectrum of monolayer MoS$_2$/few-layer graphene relative to the SiO$_2$/Si substrate. The arrows indicate contrast peaks at the A and B exciton resonances. (b) Dependence of junction photocurrent on photon energy. The arrows indicate the A and B exciton resonances. (c) Schematic of the opto-valleytronic spin injection experiment. (d) Electrical spin signal $V_{NL}$ as a function of $B_y$ exhibits clear antisymmetric Hanle spin precession signals which flip polarity with the Co magnetization direction (M+ vs. M-). The photon energy is tuned to the A exciton resonance at E = 1.93 eV.

Next, we determine the appropriate photon energy for optical spin injection into MoS$_2$ by performing optical reflection spectroscopy and photocurrent spectroscopy of the MoS$_2$/graphene heterostructure at 10 K. A focused beam (~2 μm, 100 μW) from a tunable laser is incident on the MoS$_2$/graphene heterostructure, and the reflection contrast ΔR/R (compared to the substrate) is measured as a function of incident photon energy. At the same time, the photocurrent response ($I_{ph}$) is measured across electrodes

G1 and G2. The reflection contrast spectrum (Figure 3a) shows the maximum contrast at ~1.93 eV and ~2.06 eV, which correspond to the A and B exciton resonances of monolayer $MoS_2$[6, 12, 15, 21]. Similarly, we observe two peaks at nearly identical photon energies in the photocurrent spectrum (Figure 3b) (see SI, Sec. 3 for details of the measurements).

Having established the optimal energy for light absorption and the ability to detect spins electrically, we turn our attention to the combined functionality of optical spin injection and lateral spin transport in the $MoS_2$/graphene hybrid spin valve. As illustrated in Figure 3c, we focus the laser beam (~2 μm, 100 μW) on the $MoS_2$/graphene junction at a photon energy of 1.93 eV (A exciton) for the optical spin injection, and measure the voltage $V_{NL}$ across electrodes C6 and G2 for electrical spin detection. We also magnetize the detector electrode magnetization along +x direction (denoted as M+). Circular polarization of the incident light produces spin/valley polarization in the $MoS_2$ layer with spin oriented out-of-plane (for noise rejection, we modulate the helicity and detect using lock-in techniques, as discussed in the SI, Sec. 4). A coherent transfer of spin across the $MoS_2$/graphene interface results in out-of-plane spin polarization in the graphene layer, which will subsequently diffuse towards the FM detector (C6). Because the spin orientation is perpendicular to the detector magnetization, this will result in zero spin signal in $V_{NL}$. In order to detect spins, we therefore apply an external in-plane field ($B_y$) along the graphene strip to induce spin precession and generate a component of spin along the detector magnetization (+x direction). The light blue curve in Figure 3d shows the measured voltage $V_{NL}$ as a function of $B_y$. At low fields, $V_{NL}$ varies approximately linearly with $B_y$ because the spin precession angle varies linearly with the field. At higher fields, the increase of $V_{NL}$ with $B_y$ eventually reaches a maximum and reduces as the average precession angle exceeds ~90°. Later, we provide a quantitative description of this curve, known as an "antisymmetric Hanle curve", by modeling the spin transport and precession using one-dimensional drift-diffusion equations. To verify that the signal indeed comes from spin, we reverse the FM magnetization direction (M-) and repeat the measurement. The result is an inverted $V_{NL}$ signal, as shown in the dark grey curve of Figure 3d, which is the expected behavior for a $V_{NL}$ signal generated by spin polarization. Observation of antisymmetric Hanle curves that flip with the magnetization state (M+ vs. M-) provide proof of optical spin injection into $MoS_2$, followed by coherent spin transfer across the $MoS_2$/graphene interface, lateral spin transport in graphene, and electrical spin detection. Additional measurements show that the spin transfer from $MoS_2$ to graphene is dominated by hole transport (SI, Sec. 5).

Tuning the photon energy from the A exciton to the B exciton should switch the orientation of the injected spin polarization due to the large spin-orbit splitting in the monolayer TMD band structure. As shown in Figure 4a, the valence band of monolayer $MoS_2$ has a large spin splitting with opposite spin orientation for the A and B optical excitations within the same valley. Figure 4b shows antisymmetric Hanle curves for four different photon energies: 1.87 eV, 1.93 eV, 1.96 eV, and 2.06 eV. At each photon energy, $V_{NL}$ vs $B_y$ is measured for both FM magnetization directions (M+ and M−), and we plot the subtracted signal $V_{NL,total} = V_{NL,M+} - V_{NL,M-}$ which helps cancel background signals unrelated to spin. For the A exciton resonance energy (1.93 eV), $V_{NL,total}$ has a minimum value near $B_y$ = -50 mT and increases to a maximum signal at around $B_y$ = 50 mT. As indicated in Figure 4b, we define the spin signal as $\Delta V_{NL} = V_{NL,total}(B_y= 50\text{ mT}) - V_{NL,total}(B_y= -50\text{ mT})$. Away from the A exciton resonance, the Hanle curves for 1.87 eV (black) and 1.96 eV (purple) have smaller spin signals $\Delta V_{NL}$ than on resonance. In contrast, near the B resonance (2.06 eV), the spin signal completely reverses sign to give a flipped antisymmetric Hanle curve. This indicates a reversal of spin orientation as the photon energy is tuned from the A resonance to the B resonance.

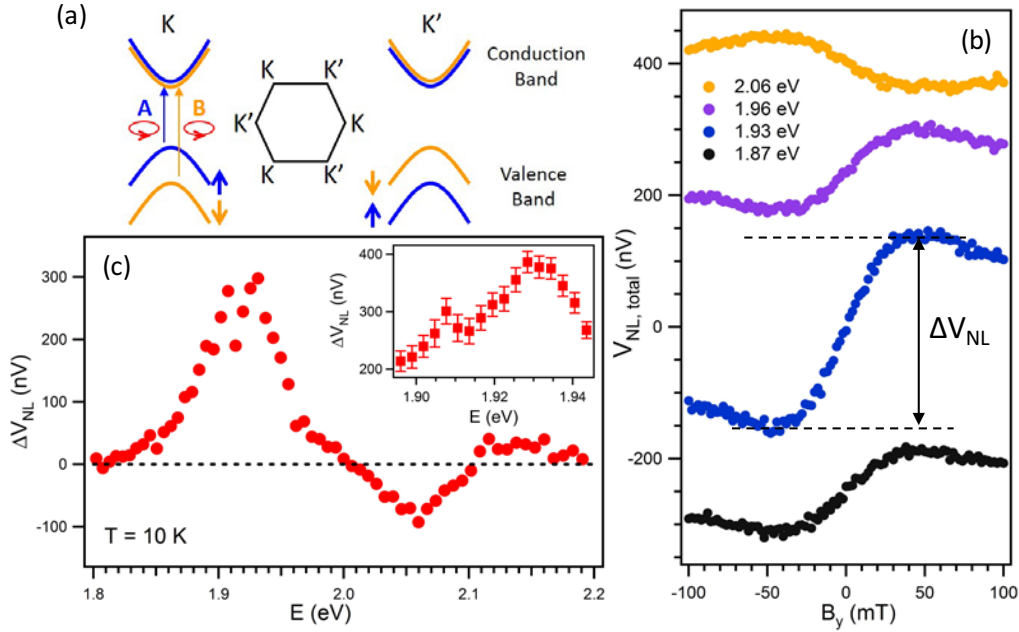

**Figure 4.** Photon energy dependence of opto-valleytronic spin injection. (a) Schematic band structure of monolayer $MoS_2$ at the K and K' valleys. (b) Representative antisymmetric Hanle curves at four photon energies (1.87 eV, 1.93 eV, 1.96 eV and 2.06 eV). (c) Spin signal $\Delta V_{NL}$ as a function of photon energy. Inset shows zoom-in detailed features around the A exciton resonance.

To further investigate the role of photon energy, we map out the detailed photon energy dependence of the spin signal $\Delta V_{NL}$ at 10 K. We obtain $\Delta V_{NL}$ at each photon energy by measuring $V_{NL,total}$ at +50 mT and -50 mT and plot the resulting $\Delta V_{NL}$ vs photon energy in Figure 4c. Starting from low photon energy, $\Delta V_{NL}$ reaches a maximum positive signal near the A resonance (1.90 - 1.95 eV), then decreases with increasing photon energy until $\Delta V_{NL}$ flips sign around 2 eV. $\Delta V_{NL}$ reaches a minimum near the B resonance at ~2.06 eV. This photon energy dependence clearly reflects the non-degenerate spin-split structure of the valence band, which results from strong spin-orbit coupling and the broken inversion symmetry of the monolayer $MoS_2$ lattice. In the vicinity of the A resonance, we observe a double peak feature. To exclude potential artifacts from noise or sample drift, we retake the data with smaller energy steps and perform a spatial mapping of the spin signal $\Delta V_{NL}$ over the $MoS_2$/graphene junction at each energy (details in SI, Sec. 6). The inset of Figure 4c plots the maximum $\Delta V_{NL}$ from the spatial map as a function of photon energy. The presence of two separate peaks near the A resonance can be clearly distinguished. The two peaks are at 1.91 eV and 1.93-1.94 eV, corresponding to a splitting of 20-30 meV. This is consistent with the double peak structure of the $A^-$ trion and A exciton, which exhibits a splitting of 20-40 meV in photoluminescence and optical absorption measurements[15, 22, 23].

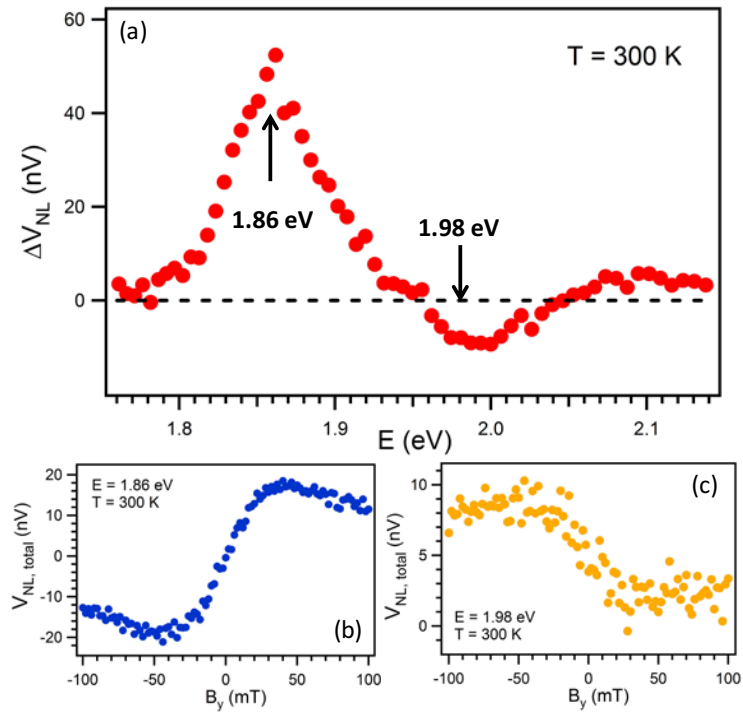

**Figure 5.** Room temperature characteristics of opto-valleytronic spin injection. (a) Photon energy dependence of the spin signal $\Delta V_{NL}$. (b) Antisymmetric Hanle curve at the A resonance (photon energy of 1.86 eV). (c) Antisymmetric Hanle curve at the B resonance (photon energy of 1.98 eV).

In addition, we explore the temperatures at which the opto-spintronic device can successfully operate. Remarkably, the signal persists up to room temperature. As shown in Figure 5a, $\Delta V_{NL}$ at room temperature exhibits a similar dependence on photon energy as at low temperature, with the positive peak at the A resonance red-shifted to around 1.86 eV, and the negative peak at the B resonance red-shifted to around 1.99 eV. The red shift and peak positions at room temperature are consistent with previous experimental and theoretical studies[24, 25]. Two Hanle scans with photon energies near the A and B peaks are measured (Figure 5b and 5c) to confirm room temperature spin orientation switching from the A resonance to the B resonance. The room temperature signal is about 5 times smaller than at 10 K. We consider various factors that can give rise to the reduced spin signal at room temperature. In standard graphene spin valves, the spin lifetime and spin diffusion length have a weak temperature dependence[26], so the graphene alone could not explain the strong temperature dependence that we observe. However, the spin and valley dependent properties in $MoS_2$ are strongly degraded with increasing temperature. As the temperature increases, there is more intervalley scattering which reduces the valley polarization[13, 27-30]. In addition, the spin lifetime of resident carriers in $MoS_2$ are also strongly reduced with increasing temperature[31, 32]. Thus, the presence of spin signal at room temperature suggests a rapid transfer of spin-polarized carriers from $MoS_2$ to graphene. Despite the smaller signal, successful room temperature operation lays the foundation for multifunctional opto-spintronic and opto-valleytronic devices in 2D materials and heterostructures. The data has been reproduced on a second sample (SI, sec. 8), and we observe similar effects in preliminary measurements on monolayer $MoS_2$/monolayer graphene samples.

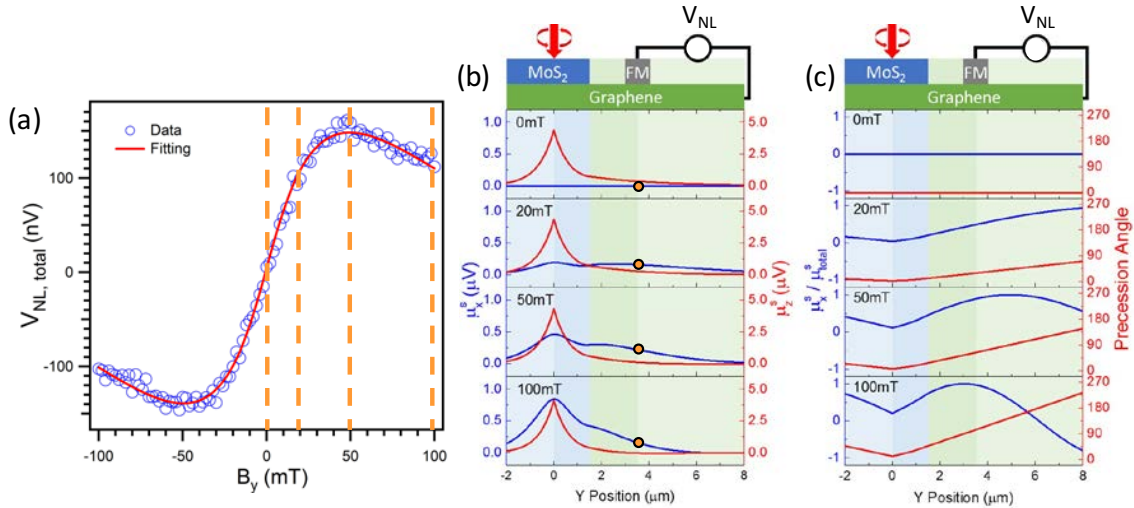

**Figure 6.** Modeling the antisymmetric Hanle curve. (a) Data (open circles) and fitting (red curve) of the antisymmetric Hanle curve for photon energy of 1.93 eV and temperature of 10 K. (b) The x and z components of spin accumulation in the few-layer graphene as a function of position. The orange circles represent the component of spin accumulation ($\mu_x^s$) measured by the non-local voltage $V_{NL}$. (c) $\mu_x^s/\mu_{total}^s$ and the average precession angle as a function of position.

To better understand the data and quantify the optical spin injection current, we have developed a one-dimensional model to describe spin transport in the monolayer $MoS_2$/few-layer graphene hybrid spin valve. In our model, spin accumulation is considered as a three-component vector ($\mu_x^s$, $\mu_y^s$, $\mu_z^s$), where each component describes the spin polarization in different directions. The optically injected spin current from MoS$_2$ to graphene is modeled as a point source at the center of the laser spot, and we assume that the MoS$_2$/graphene and pure graphene regions have different spin transport parameters due to the additional spin relaxation induced by the MoS$_2$. In addition, the MoS$_2$ could generate proximity-induced spin-orbit coupling in the graphene.[20, 33, 34] However, such effects are not apparent in our data and therefore not incorporated in our model. The lateral spin transport and spin precession are modeled using the steady-state Bloch equation. Details of the model are provided in SI, Sec. 9, with Figure 6a showing the best fit to the experimental data taken from Figure 4b (1.93 eV curve). The spin lifetime $\tau_G$ = 308 ps and spin diffusion coefficient $D_G$ = 0.0301 m$^2$/s from the electrical non-local Hanle measurements (Figure 2d) are used as fixed parameters for the pure graphene region, while $\tau_M$ and $D_M$ for the MoS$_2$/graphene region are fitting parameters. The best fit yields $\tau_M$ = 23.9 ps and $D_M$ = 0.0183 m$^2$/s in the MoS$_2$/graphene region and an optically injected spin current of 116 nA.

The corresponding spatial profiles of the spin accumulation components $\mu_x^s$ and $\mu_z^s$ are shown in Figure 6b for representative fields $B_y$ = 0 mT, 20 mT, 50 mT, and 100 mT. Because the magnetization of the spin detector is along the x-axis, the measured signal is proportional to $\mu_x^s$ at y = L (detector position) as indicated by the orange circles in Figure 6b. For $B_y$ = 0 mT, the spin population in the channel diffuses without precession. Thus, $\mu_x^s$ = 0 is zero throughout the channel, leading to $V_{NL}$ = 0 at the detector. As the magnetic field is turned on, the spins start to precess while diffusing, and the x-component of spin accumulation begins to build up in the channel. At $B_y$ = 50 mT, the x component of spin accumulation underneath the contact reaches a maximum, which results in a maximum non-local voltage in Figure 6a. The spin precession is best illustrated through the spatial profiles of $\mu_x^s/\mu_{total}^s$ which is the unit vector of spin accumulation projected along the detector magnetization, as shown in Figure 6c. The average precession angle is given as arctan($\mu_x^s/\mu_{total}^s$). With increasing magnetic field, the precession angle vs. position increases in slope as expected, and the position for maximum $\mu_x^s/\mu_{total}^s$ moves to the left, closer to the source point. As this maxima passes by the detector, any further increase of $B_y$ leads to a reduced spin signal, which explains why the antisymmetric Hanle curve in Figure 6a decreases for $B_y$ > 50 mT.

In conclusion, we demonstrate opto-valleytronic spin injection in monolayer MoS$_2$/few-layer graphene hybrid spin valves through Hanle spin precession measurements. The magnitude and direction of optically injected spins are tunable by both helicity and photon energy, and the observed spin signals persist up to room temperature. In terms of scaling, such opto-spintronic devices would be subject to the diffraction limit (~500 nm), although the use of near-field optics could allow for smaller devices. These results pave the way for multifunctional 2D spintronic/valleytronic devices and applications.

## ASSOCIATED CONTENT

**Supporting Information**

A description of monolayer $MoS_2$/few-layer graphene hybrid spin valve device fabrication, graphene spin transport measurement and analysis, optical reflection and photocurrent spectroscopy (charge currents), experimental setup for the optical injection and electrical detection of spin currents, identifying the carrier type for opto-valleytronic spin injection, spatial mapping of the spin signal, additional measurements, and details of the modeling.

## AUTHOR INFORMATION

**Corresponding Author**

*E-mail: kawakami.15@osu.edu

**Author Contributions**

R. K. K. conceived the project. Y. K. L. and J. X. fabricated the devices and performed the measurements. G. W., E. J. M. and W. Z. assisted in the measurements. T. Z. and M. R. N. contributed to the modeling and theoretical analysis. All authors contributed to writing the manuscript.

## ACKNOWLEDGEMENTS

We acknowledge helpful discussions with Vladimir Falko and technical assistance from Jyoti Katoch and Simranjeet Singh. This work was primarily supported by NSF DMR-1310661 (optical measurements). We acknowledge partial support from the Center for Emergent Materials: an NSF MRSEC under award number DMR-1420451 (sample fabrication).

For TOC only:

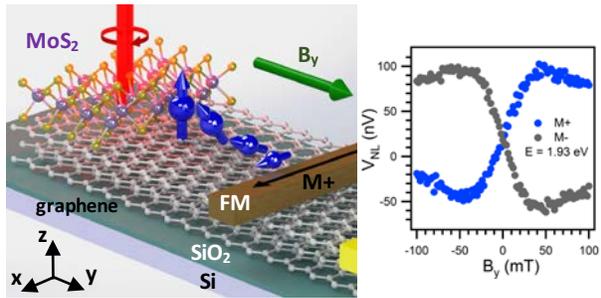

**Supporting Information for:**

**Opto-Valleytronic Spin Injection in Monolayer MoS$_2$/Few-Layer Graphene Hybrid Spin Valves**

Yunqiu (Kelly) Luo,[1,#] Jinsong Xu,[1,#] Tiancong Zhu,[1] Guanzhong Wu,[1] Elizabeth J. McCormick,[1] Wenbo Zhan,[1] Mahesh R. Neupane,[2] Roland K. Kawakami[1,*]

[1]*Department of Physics, The Ohio State University, Columbus, OH 43210, USA*

[2]*Sensors and Electron Devices Directorate, U.S. Army Research Laboratory, Adelphi, Maryland 20783, USA*

[#]equal contributions

*Corresponding Author

   e-mail:     kawakami.15@osu.edu

   Address:   191 W. Woodruff Ave.
                   Department of Physics
                   The Ohio State University
                   Columbus, OH 43210

   Phone:     (614) 292-2515

   Fax:       (614) 292-7557

# 1. Fabrication of Monolayer MoS$_2$/Few-Layer Graphene Hybrid Spin Valves

The monolayer MoS$_2$/few-layer graphene heterostructures are prepared by first exfoliating a few-layer graphene flake onto heavily n-doped Si substrates with a 300 nm SiO$_2$ layer. Monolayer MoS$_2$ is then transferred onto the few-layer graphene. For the transfer, we mount ~2 mm thick polydimethylsiloxane (PDMS) on a glass slide and cover it with a thin film of polycarbonate (PC). This PC/PDMS stamp is used to pick up an exfoliated monolayer MoS$_2$ flake from an SiO$_2$/Si substrate. The MoS$_2$ flake is then aligned and brought into contact with the few-layer graphene on an SiO$_2$/Si substrate. After contact, the PC film is cut from the glass slide and the entire PC/MoS$_2$ combination remains on the few-layer graphene/SiO$_2$/Si substrate. The PC film is then dissolved in chloroform. After that, the transferred graphene/MoS$_2$ heterostructure is cleaned of polymer residue by annealing at 350 °C in ultra-high vacuum (UHV) for 1 hour. Figure S1a shows the sample after this step. It is important to note that while polymers are used in the transfer process, the surfaces that form the MoS$_2$/graphene interface are never in contact with the polymers. Next, the MoS$_2$/graphene heterostructure is patterned by e-beam lithography with PMMA resist and etched by low-power inductively coupled plasma reactive ion etch (ICP-RIE) to have two (or more) graphene strips extend from the junction region. This process is followed by another annealing step in UHV to remove PMMA residue. Figure S1b shows the sample after this step. Subsequently, we use two steps of e-beam lithography with MMA/PMMA bilayer resist to fabricate electrodes. In the first step, Au electrodes (70 nm) are deposited on the few-layer graphene using an e-beam source and a 5 nm Cr underlayer for adhesion (Figure S1c). In the second step, Co electrodes with SrO tunnel barriers are deposited in an MBE chamber using angle evaporation with polar angle of 0° for the SrO masking layer (3 nm), 9° for the SrO tunnel barrier (0.8 nm), and 7° for the Co electrode (60 nm)[1,2]. Figure S1d shows an image of the completed device.

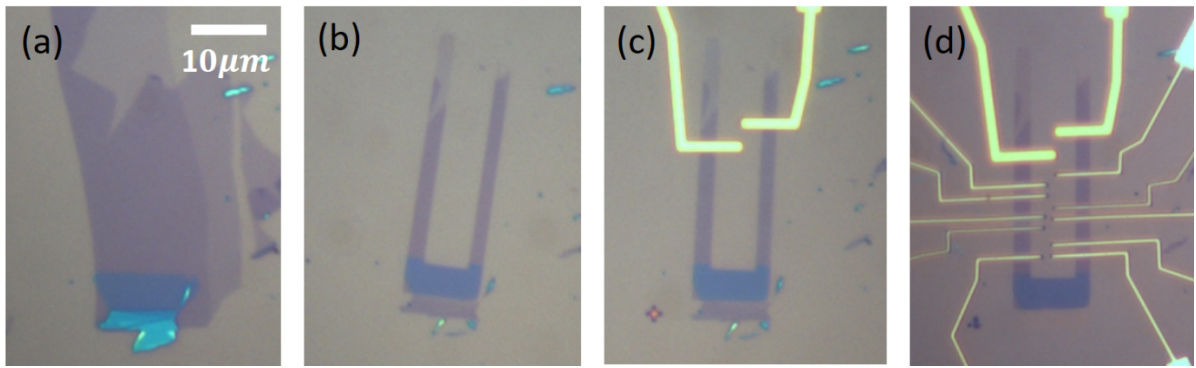

**Figure S1.** Device fabrication process: optical images of MoS$_2$/graphene heterostructure (a) after UHV annealing, (b) after ICP-RIE etch, (c) after Au electrode deposition, and (d) after deposition of SrO tunnel barrier and Co ferromagnetic electrode.

Figure S2 shows additional characterizations of the material. First, the thickness of the few-layer graphene is characterized by Raman spectroscopy and atomic force microscopy (AFM). The Raman spectrum of the 2D peak (Figure S2a) is consistent with a thickness of three or more layers,[3] while AFM measurements indicate a thickness of 3-4 layers.[4] Second, the few-layer graphene is determined to be n-type by measuring the four-probe resistance as a function of backgate voltage $V_G$ applied to the Si substrate (Figure S2b). The decrease of resistance with increasing $V_G$ indicates that the few-layer graphene is n-type. Unless specifically noted, all measurements are performed at $V_G = 0$ V.

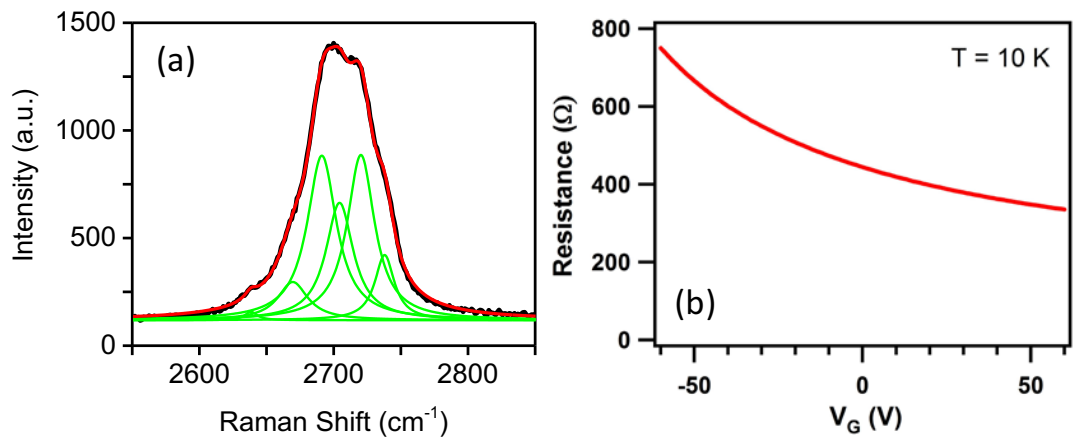

**Figure S2.** Characterization of the few-layer graphene. (a) Raman spectrum of the 2D peak, which indicates a graphene thickness of three or more layers (514 nm laser wavelength). (b) Gate-dependent resistance, indicating that the graphene is n-type.

## 2. Details of the Graphene Spin Transport Measurement and Analysis

Here we provide details of the spin transport measurement as shown in Figure 2 of the main text. The device is wired up in a non-local spin transport geometry as shown in Figure 2b, and we utilize lock-in detection with a modulation frequency of 11 Hz for noise rejection. Electrical spin injection into graphene is performed by passing an AC current of 1 µA rms ($I_{inj}$) between the electrodes C6 (spin injector) and G1, and the spin transport signal is measured as an AC modulated non-local voltage ($V_{NL}$) between electrodes C8 (spin detector) and G2 (nonmagnetic reference electrode). Electrodes C6 and C8 have contact resistances of 3.4 kΩ and 65 kΩ, respectively, due to the SrO tunnel barriers, while electrodes G1 and G2 are ohmic contacts. We use a Stanford Research 560 voltage preamplifier followed by a Signal Recovery 7265 lock-in amplifier for detection of $V_{NL}$. To separate spin transport signals from charge-related backgrounds, we sweep an external magnetic field $B_x$ parallel to the Co electrodes while recording $V_{NL}$. Figure 2c shows the observed $V_{NL}$ as a function of $B_x$, where the red (blue) curve is for the sweep with increasing (decreasing) value of $B_x$. For the red curve, we see an abrupt change in $V_{NL}$ at ~25 mT, where one of the Co electrode magnetizations flips to create an antiparallel alignment of injector C6 and detector C8. With $B_x$ further increased (~28 mT), $V_{NL}$ changes back to its original value as both injector and detector flip to parallel alignment. Similar behavior when decreasing $B_x$ (blue curve) is also observed. The observed jumps in $V_{NL}$ are the hallmark of electrical spin transport, whereas the overall background is unrelated to spin. We note the presence of an additional Co electrode C7 that lies between the spin injector C6 and spin detector C8. Because C7 has a large contact resistance of ~100 kΩ, we assume it has a negligible effect on spin transport between contacts C6 and C8.

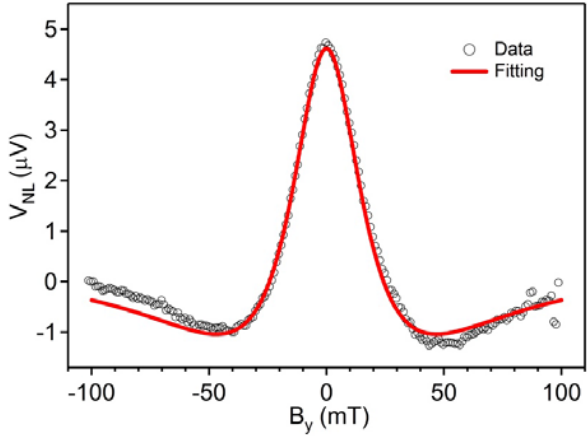

**Figure S3.** In-plane Hanle curve measured at T = 10 K. The black circles are the measured data and the red curve is the best fit.

To determine the spin relaxation time, diffusion coefficient and spin diffusion length in the few-layer graphene channel, we perform in-plane Hanle spin precession measurements in the non-local geometry. An external in-plane magnetic field ($B_y$) perpendicular to the Co electrode magnetization is applied to induce spin precession. Figure 2d shows the Hanle curves obtained for the graphene channel measured in parallel (red curve) and antiparallel (blue curve) alignments of the injector and detector magnetizations. The measured sweeps for parallel and antiparallel configurations are subtracted and the spin relaxation time, diffusion coefficient, and spin diffusion length are determined by fitting to an analytical expression developed by Sosenko *et al.*[5]. For the fit, we use the measured contact resistances of the injector and detector to be 3.4 kΩ and 65 kΩ, respectively, and the sheet resistance of the graphene channel to be 340 Ω. Figure S3 shows best fit curve (red) overlaid on the subtracted data (black dots), which yields a spin relaxation time of $\tau_G$ = 308 ± 13 ps, a diffusion coefficient of $D_G$ = 0.0301 ± 0.0013 m²/s, and an effective spin polarization of the electrodes $P_{eff}$ = 0.317 ± 0.017. The corresponding spin diffusion length is $\lambda_G$ = $\sqrt{D_G \tau_G}$ = 3.04 μm. We notice the presence of asymmetry in the Hanle curve, so we repeated the analysis on just the symmetric component, $V_{NL}^s(B_y) = [V_{NL}(B_y) + V_{NL}(-B_y)]/2$, and obtained $\tau_G$ = 308 ± 5 ps and $D_G$ = 0.0301 ± 0.0006 m²/s. This indicates that the effect of the asymmetry is negligible.

# 3. Optical Reflection and Photocurrent Spectroscopy (Charge Currents)

To perform the optical reflection spectroscopy and photocurrent spectroscopy shown in Figures 3a and 3b of the main text, respectively, we use a tunable laser source (Fianium Supercontinuum WhiteLase and LLTF) coupled with high-precision XYZ scanning stages (Newport) and an ultra-low vibration optical cryostat (Advanced Research Systems). Figure S4a is a schematic of the optical setup. The sample is held fixed inside the optical cryostat and a 50x objective (Mitutoyo) focuses the laser beam down to a ~2 µm diameter spot. The objective and mirrors M1 and M2 are mounted on the XYZ stage to allow two-dimensional scanning of the sample. A non-polarizing beamsplitter (BS) is mounted before M1 to direct the reflected beam coming from sample into a photodiode detector. An optical chopper (not drawn in Figure S4a) is mounted earlier in the beam path to modulate the intensity of the laser beam for lock-in detection.

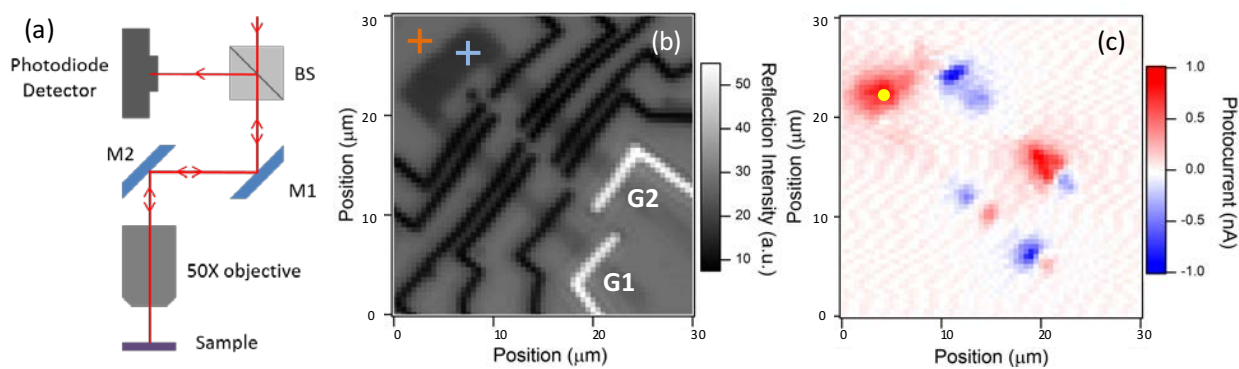

**Figure S4.** (a) Optical setup for reflectivity and photocurrent microscopy. (b) Reflectivity image of the device shown in Figure 2a of the main text. The blue crosshair is the laser position for measuring R′, and the orange crosshair is the position for measuring R. (c) Photocurrent map taken simultaneously with the reflectivity image. The yellow dot is the laser position with maximum photocurrent response from the $MoS_2$/graphene junction.

First, we image the device by monitoring the reflectivity at a fixed wavelength (600 nm) and constant power of 30 µW. By scanning the focused laser spot on the device area while measuring the reflected beam intensity, a real space image of the device is obtained, as shown in Figure S4b. To measure the optical reflection contrast spectrum of the monolayer $MoS_2$/few-layer graphene junction, the laser power is increased to 100 µW. The reflection intensity R′ is measured as a function of laser wavelength λ at the junction position (blue crosshair in Figure S4b) to obtain R′(λ) and at a nearby position on the substrate (orange crosshair in Figure S4b) to obtain R(λ). The reflection contrast spectrum shown in Figure 3a of the main text is given by ΔR/R, where ΔR = R − R′.

In addition, during the spatial mapping of the device by reflectivity (Figure S4b), we simultaneously measure the photocurrent response by monitoring the charge current using electrodes G1 and G2 (G2 grounded, G1 goes into SR570 then lock-in). As shown in a spatial map of the photocurrent in Figure S4c, there is a strong photocurrent when the laser beam is on the $MoS_2$/graphene junction and when it is near the metallic electrodes. The positive photocurrent on the $MoS_2$/graphene junction indicates that photocurrent generated at the heterostructure flows towards G1. To perform photocurrent spectroscopy of the $MoS_2$/graphene junction, we move the laser beam to the position of highest response (yellow dot in Figure S4c) and measure the photocurrent as a function of the photon energy (at constant power of 100 µW). The resulting data is plotted in Figure 3b of the main text.

## 4. Experimental Setup for the Optical Injection and Electrical Detection of Spin Currents

For the optical injection and electrical detection of the spin current presented in Figures 3d, 4b-c, and 5a-c of the main text, we insert additional polarization optics into the previously described scanning reflectivity and photocurrent microscopy setup (section 3). For these measurements, we utilize an average power of 100 µW. As illustrated in Figure S5, a linear polarizer (LP) is inserted in the beam path between the non-polarizing beamsplitter (BS) and M1. Between M1 and M2, we insert a liquid crystal variable retarder (LCVR) (Meadowlark) to modulate the polarization of the laser beam between $+\lambda/4$ (RCP) and $-\lambda/4$ (LCP) at a frequency of $f = 11$ Hz. Electrical detection of the spin signal is performed by measuring the non-local voltage $V_{NL}$ (see Figure 3c of the main text) using a voltage preamplifier (Stanford Research 560) and lock-in amplifier (Signal Recovery 7265). By using helicity modulation instead of intensity modulation, we can effectively suppress the photocurrent background and therefore detect the signal that is only sensitive to the polarization state of the incident beam.

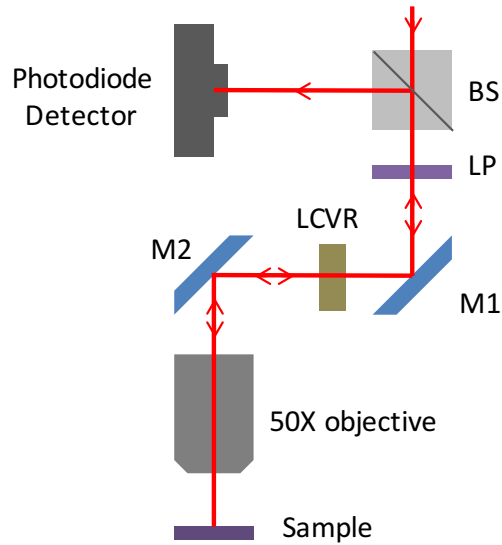

**Figure S5.** Optical setup for optical spin injection and electrical spin detection. Two extra polarization optics, the liquid crystal variable retarder (LCVR) and linear polarizer (LP), are added to the setup in Figure S4a.

# 5. Identifying the Carrier Type for Opto-Valleytronic Spin Injection

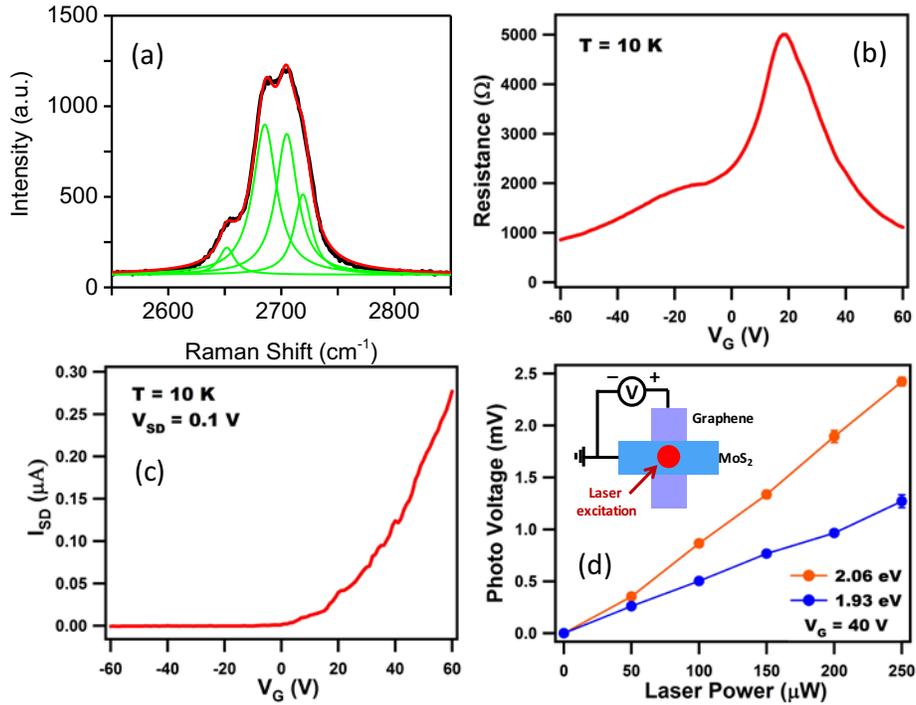

**Figure S6.** Photovoltage measurement of a monolayer MoS$_2$/few-layer graphene junction, performed at 10 K. (a) Raman spectrum of the graphene region indicates bilayer thickness (514 nm laser wavelength). (b) Gate-dependent resistance of the graphene. (c) Gate-dependent source-drain current ($I_{SD}$) at fixed bias voltage ($V_{SD}$ = 0.1 V) of the MoS$_2$. (d) Photovoltage as a function of laser power for photon energies of 1.93 eV (A resonance) and 2.06 eV (B resonance). Inset: A diagram showing the polarity of the DC voltage measurement, with the graphene connected to the positive terminal and the MoS$_2$ connected to the negative terminal.

To determine whether the spin transfer from monolayer MoS$_2$ to few-layer graphene is dominated by electron or hole transport, we perform photovoltage measurements on a monolayer MoS$_2$/few-layer graphene junction in a crossbar geometry (inset of Figure S6d) on SiO$_2$(300 nm)/Si substrate (used as backgate). The graphene thickness is determined to be bilayer based on Raman spectroscopy (Figure S6a). Four-probe gate-dependent resistance measurements of the graphene and gate-dependent source-drain current ($I_{SD}$) at fixed bias voltage ($V_{SD}$ = 0.1 V) of the MoS$_2$ are shown in Figures S6b and S6c, respectively. For the photovoltage study, we set the backgate voltage to $V_G$ = +40 V for n-type graphene because the opto-valleytronic spin injection experiments utilize n-type graphene (see Figures S2 and S9). The voltage signal is amplified by a Stanford 560 pre-amp and measured by a DC voltmeter (Keithley 2002) with the graphene connected to the positive terminal and the MoS$_2$ connected to the negative terminal (inset of Figure S6d). Figure S6d shows the DC photovoltage vs. laser power when the junction is illuminated with either 1.93 eV photons (A resonance) or 2.06 eV photons (B resonance). In both cases, the illumination of the junction produces a positive photovoltage, which indicates that the transfer of carriers from the monolayer $MoS_2$ to the few-layer graphene is dominated by hole transport. Therefore, the opto-valleytronic spin injection is dominated by the transfer of spin-polarized holes from $MoS_2$ to graphene. We note that the same polarity of photovoltage is also observed for measurements performed with $V_G$ = 0 V and $V_G$ = -30 V, which shows that the charge transport from $MoS_2$ to graphene is dominated by holes even when the graphene is p-type.

## 6. Spatial Mapping of the Spin Signal

To verify the double peak structure of the spin signal $\Delta V_{NL}$ vs. photon energy observed near the A exciton resonance (Figure 4c of the main text), we perform more detailed scans to rule out potential artifacts from noise or sample drift. To accomplish this, we obtain spatial maps (6 μm × 6 μm) of the spin signal $\Delta V_{NL}$ on the $MoS_2$/graphene junction and at finer energy steps (3 meV). Figure S7 (right column) shows the spin signal mapping at two representative photon energies (1.925 eV and 1.937 eV). While mapping the spin signal, we simultaneously map the reflectivity to track the sample position (left column in Figure S7). The red dashed lines in Figure S7 outline the boundaries of the $MoS_2$/graphene junction. Due to the insertion of polarization optics in Figure S5, the reflected intensity will be modulated at a frequency of $2f$. This occurs because the LP and LCVR are in a Faraday isolator geometry so that retardance of + λ/4 (RCP) and - λ/4 (LCP) both yield a minimum reflection intensity, whereas a zero retardance for the LCVR produces a maximum reflection intensity. Comparing the maps of the spin signal and reflectivity at different wavelengths, we observe that the maximum spin signal occurs at the same sample position within the microscope resolution. From this data set, we extract the maximum spin signal for each photon energy, and the results are presented in the inset of Figure 4c in the main text. The double peak feature around the A exciton resonance is clearly observed within the error bars of the measurement, which confirms the initial observation reported in the main panel of Figure 4c. The error bars reflect the standard deviation of the measurement of $\Delta V_{NL}$, which is obtained by repeated measurements of $V_{NL}$ for $B_y$= 50 mT and $B_y$= -50 mT (and for M+ and M- magnetizations).

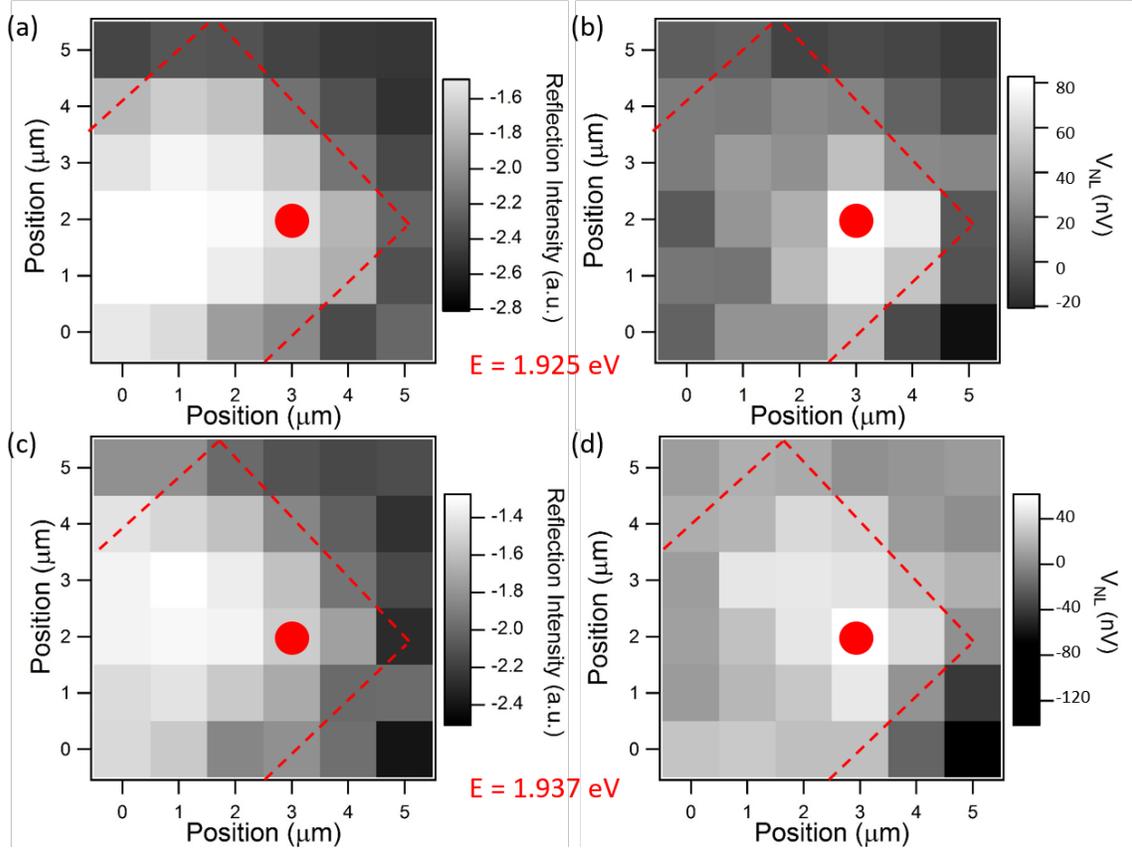

**Figure S7.** (a) Reflectivity image for photon energy of 1.925 eV. (b) Simultaneous spin signal $V_{NL}$ mapping for 1.925 eV. (c) Reflectivity image for photon energy of 1.937 eV. (d) Simultaneous spin signal $V_{NL}$ mapping for 1.937 eV. In all maps, the red dot indicates the largest signal and the red dashed lines outline the boundary of $MoS_2$/graphene junction.

## 7. Laser Power Dependence of the Spin Signals

To study the laser power dependence of the spin signal, we measure $\Delta V_{NL}$ at both A (1.93 eV photon energy) and B (2.06 eV photon energy) exciton resonances ($\Delta V_{NL} = V_{NL,total}(B_y = 50\ mT) - V_{NL,total}(B_y = -50\ mT)$, where $V_{NL,total} = V_{NL,M+} - V_{NL,M-}$). As illustrated in Figure S8, a linear dependence is observed in the power range of 0 μW to 110 μW.

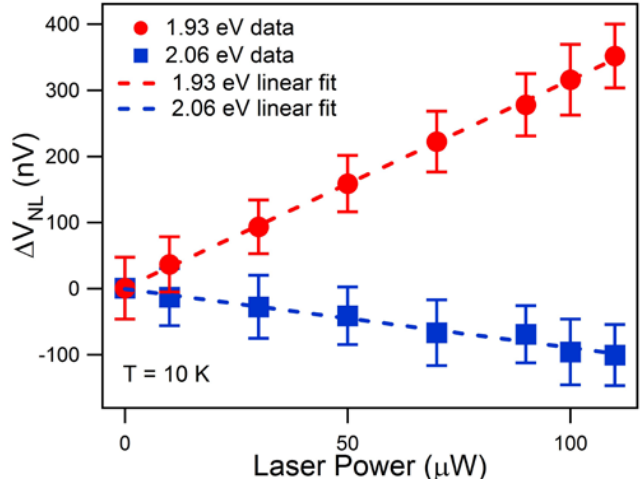

**Figure S8.** Laser power dependence of the spin signal $\Delta V_{NL}$ at A (1.93 eV photon energy, red circles) and B (2.06 eV photon energy, blue squares) exciton resonances. Dashed lines are linear fits for each photon energy.

## 8. Additional Data

We also report the result of experiments performed on another sample, which we will call "sample 2" (the sample in the main text will henceforth be called "sample 1"). As indicated by Raman spectroscopy and gate-dependent resistance (Figure S9), the graphene is n-type and has a thickness of 2 layers.[3] Unless specifically noted, all measurements are performed at $V_G = 0$ V. For sample 2, we observe a photon energy dependence of the spin signal $\Delta V_{NL}$ (Figure S10a) that is similar to sample 1 (Figure 4c of main text). At A and B resonances (1.93 eV and 2.06 eV), the antisymmetric Hanle spin signal completely reverses sign (Figure 4b and Figure S10b). In addition, fine scans of the second sample around the A resonance again reveal a double peak with energy splitting of ~20 meV (Figure S10a inset). To reduce the measurement time while preserving the important features, sample 2 is measured using the same process in the main text with the exception that $V_{NL}$ and $\Delta V_{NL}$ are taken only for the detector magnetization along +x direction (M+). Therefore, they are labeled as $V_{NL,M+}$ and $\Delta V_{NL,M+}$ in Figure S10 and S11.

In addition, room temperature spin signals are observed on sample 2. As shown in Figure S11a, the photon energy dependence of $\Delta V_{NL,M+}$ at 300 K is red shifted compared to 10 K. The positive A exciton peak is red-shifted to ~1.87 eV, and the negative B exciton peak is red-shifted to ~1.99 eV. Two field dependent Hanle scans ($V_{NL,M+}$) at 1.87 eV and 1.99 eV are measured (Figure S11b and S11c) to confirm room temperature spin orientation switching from the A resonance to the B resonance.

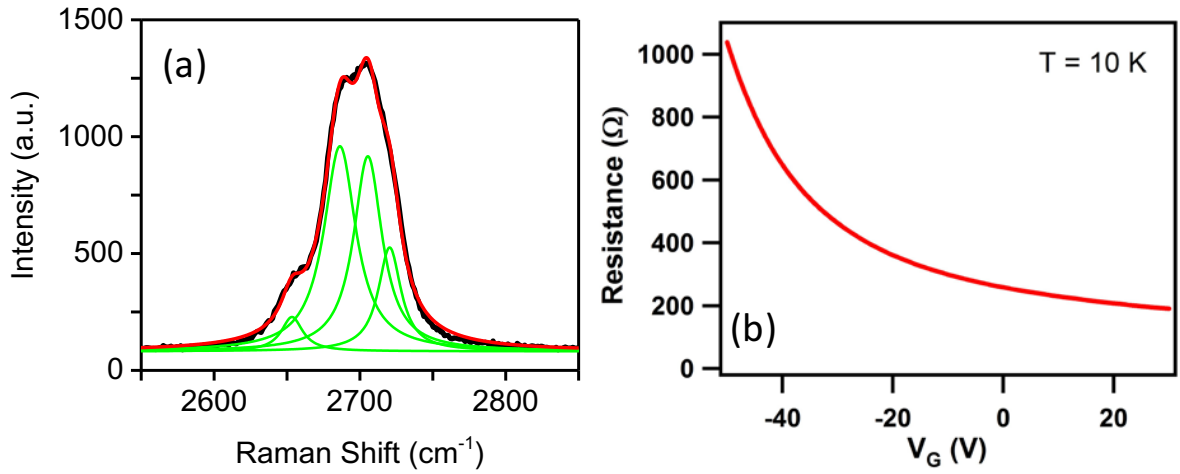

**Figure S9.** (a) Raman spectrum of the graphene region indicates bilayer thickness (514 nm laser wavelength). (b) Gate-dependent resistance of the graphene.

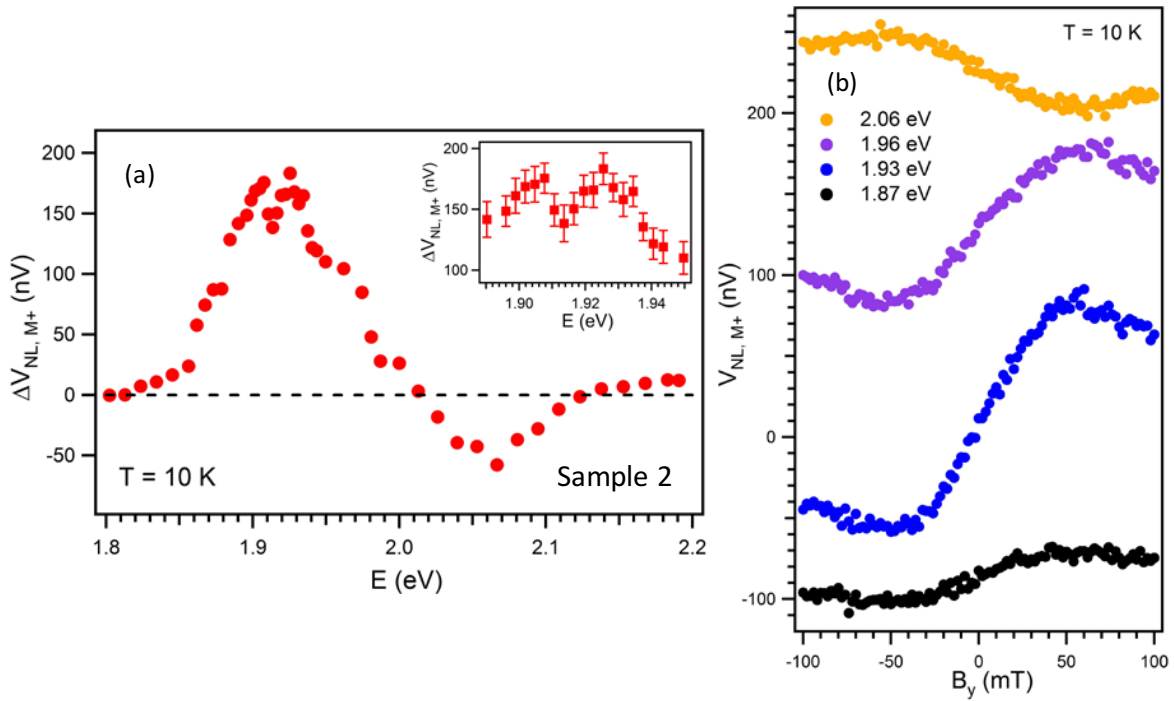

**Figure S10.** Photon energy dependence of opto-valleytronic spin injection on sample 2 with detector magnetization along +x direction (+M) at 10 K. (a) Optical Hanle signal strength. Inset shows a zoom-in of the double-peak feature around the A resonance. (b) Representative Hanle curves at four photon energies (1.87 eV, 1.93 eV, 1.96 eV and 2.06 eV).

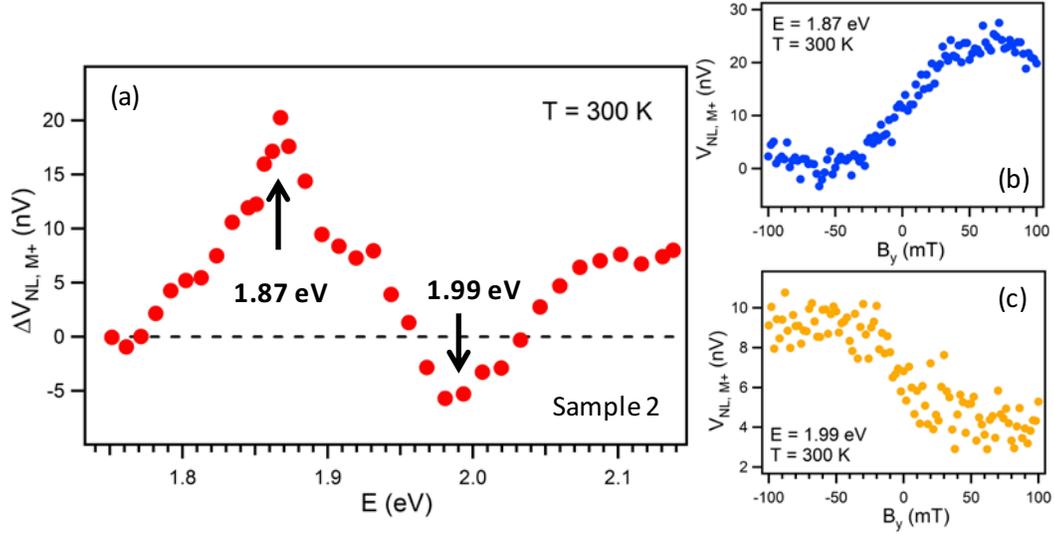

**Figure S11.** Room temperature characteristics of opto-valleytronic spin injection with detector magnetization along +x direction (M+) (a) Photon energy dependence of the spin signal $\Delta V_{NL,M+}$. (b) Antisymmetric Hanle curve at the A resonance (photon energy of 1.87 eV). (c) Antisymmetric Hanle curve at the B resonance (photon energy of 1.99 eV).

## 9. Details of the Modeling

We develop a one-dimensional model to describe the spin transport in the MoS$_2$/graphene heterostructure[5, 6]. The schematics of the measured device and device used for modeling are shown in Figure S12. In the following section, we will describe the model we use in detail.

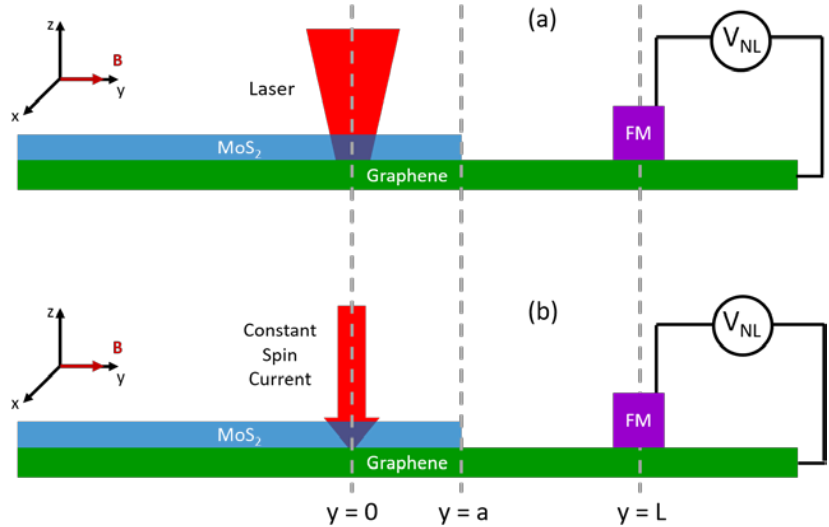

**Figure S12.** (a) Schematic of the measured device. (b) Schematic of the model discussed.

## 9.1 Optical Spin Injection into the MoS$_2$/Graphene Heterostructure

The model for spin and charge transport is one-dimensional in position (along the y-axis), and the electrochemical potential and current are spin-dependent. Because charge is a scalar and spin is a vector, there are four independent components for the spin-dependent electrochemical potential and the spin-dependent current. For the spin-dependent electrochemical potentials, we define $\vec{\mu}^\uparrow = (\mu_x^\uparrow, \mu_y^\uparrow, \mu_z^\uparrow)$ and $\vec{\mu}^\downarrow = (\mu_x^\downarrow, \mu_y^\downarrow, \mu_z^\downarrow)$, and impose the constraint that $\mu_x^\uparrow + \mu_x^\downarrow = \mu_y^\uparrow + \mu_y^\downarrow = \mu_z^\uparrow + \mu_z^\downarrow$. The average electrochemical potential for charge is defined as

$$\mu^c = \frac{\mu_x^\uparrow + \mu_x^\downarrow}{2} = \frac{\mu_y^\uparrow + \mu_y^\downarrow}{2} = \frac{\mu_z^\uparrow + \mu_z^\downarrow}{2} \tag{1}$$

and the three components of the electrochemical potential for spin are defined as

$$\vec{\mu}^s = \left(\frac{\mu_x^\uparrow - \mu_x^\downarrow}{2}, \frac{\mu_y^\uparrow - \mu_y^\downarrow}{2}, \frac{\mu_z^\uparrow - \mu_z^\downarrow}{2}\right) \tag{2}$$

For the spin-dependent currents, we define $\vec{I}^\uparrow = (I_x^\uparrow, I_y^\uparrow, I_z^\uparrow)$ and $\vec{I}^\downarrow = (I_x^\downarrow, I_y^\downarrow, I_z^\downarrow)$, and impose the constraint that $I_x^\uparrow + I_x^\downarrow = I_y^\uparrow + I_y^\downarrow = I_z^\uparrow + I_z^\downarrow$. The charge current is defined as

$$I^c = I_x^\uparrow + I_x^\downarrow = I_y^\uparrow + I_y^\downarrow = I_z^\uparrow + I_z^\downarrow \tag{3}$$

and the three components of spin current are defined as

$$\vec{I}^s = (I_x^\uparrow - I_x^\downarrow, I_y^\uparrow - I_y^\downarrow, I_z^\uparrow - I_z^\downarrow) \tag{4}$$

Due to the optical selection rules in transition metal dichalcogenides, the circularly polarized light creates a net spin accumulation polarized perpendicular to the sample surface (along the z-axis). The imbalanced spin accumulation can generate a spin current in the MoS$_2$/graphene heterostructure. This can be modeled as a pure spin current flowing into the MoS$_2$/graphene heterostructure at a constant rate

$$\vec{I}^s_{inj} = \vec{I}^\uparrow_{inj} - \vec{I}^\downarrow_{inj} = (0, 0, I^s_{z\,inj}) \tag{5}$$

$$I^c_{inj} = I^\uparrow_{inj} + I^\downarrow_{inj} = 0 \tag{6}$$

To simplify the modeling, we assume that the spin current is injected only at the center of the laser spot on the MoS$_2$/graphene heterostructure.

## 9.2 Spin Transport in the MoS$_2$/Graphene Heterostructure and Graphene Channel

Spin transport in the MoS$_2$/graphene heterostructure and graphene channel are assumed to obey the steady-state Bloch equation. The MoS$_2$/graphene heterostructure is considered as one spin transport channel, with its characteristic spin lifetime ($\tau_M$), diffusion coefficient ($D_M$) and conductivity ($\sigma_M$). The graphene channel is described with a different set of parameters ($\tau_G$, $D_G$ and $\sigma_G$, respectively). Since the MoS$_2$ resistance is normally much larger than graphene, we assume that the charge transport in the MoS$_2$/graphene heterostructure is dominated by the graphene sheet, which leads to $\sigma_M \approx \sigma_G = \sigma^c$.

For a nonmagnetic material, the charge and spin current can be calculated from the spatial distribution of the electrochemical potential

$$I^c = I^\uparrow + I^\downarrow = -w \cdot \sigma^c \nabla \mu^c \tag{7}$$

$$I^s = I^\uparrow - I^\downarrow = -w \cdot \sigma^c \nabla \mu^s. \tag{8}$$

Here $w$ is the width of the spin diffusion channel. In the device modeled, the charge current is zero throughout the whole channel. This leads to no spatial variation of $\mu^c$. We assume

$$\mu^c = 0. \tag{9}$$

To solve for the spin accumulation under an external field $\vec{B} = (0, B_y, 0)$, one can write down the steady-state Bloch equation

$$D\nabla^2 \vec{\mu}^s - \frac{\vec{\mu}^s}{\tau} + \vec{\omega} \times \vec{\mu}^s = 0, \tag{10}$$

where $D$ is the diffusion coefficient, and $\tau$ is the spin lifetime in the corresponding channel. $\vec{\omega} = (g\mu_B/\hbar)\vec{B}$ is the spin precession frequency under the external magnetic field. The above equation can be solved using Fourier transformation. By considering the boundary condition at infinity that $\lim_{x \to \pm\infty} \vec{\mu}^s = 0$, the general solution to the above equation yields

$$\mu_x^s(y) = \begin{cases} B_{-\infty} e^{\kappa_M y} + C_{-\infty} e^{\tilde{\kappa}_M y} & y < 0 \\ B_{0 \to a}^+ e^{\kappa_M y} + B_{0 \to a}^- e^{-\kappa_M y} + C_{0 \to a}^+ e^{\tilde{\kappa}_M y} + C_{0 \to a}^- e^{-\tilde{\kappa}_M y} & 0 < y < a \\ B_{a \to L}^+ e^{\kappa_G y} + B_{a \to L}^- e^{-\kappa_G y} + C_{a \to L}^+ e^{\tilde{\kappa}_G y} + C_{a \to L}^- e^{-\tilde{\kappa}_G y} & a < y < L \\ B_{+\infty} e^{-\kappa_G y} + C_{+\infty} e^{-\tilde{\kappa}_G y} & y > L \end{cases} \tag{11}$$

$$\mu_y^s(y) = \begin{cases} A_{-\infty} e^{k_M y} & y < 0 \\ A_{0 \to a}^+ e^{k_M y} + A_{0 \to a}^- e^{-k_M y} & 0 < y < a \\ A_{a \to L}^+ e^{k_G y} + A_{a \to L}^- e^{-k_G y} & a < y < L \\ A_{+\infty} e^{-k_G y} & y > L \end{cases} \tag{12}$$

$$\mu_z^s(y) = \begin{cases} iB_{-\infty} e^{\kappa_M y} - iC_{-\infty} e^{\tilde{\kappa}_M y} & y < 0 \\ iB_{0 \to a}^+ e^{\kappa_M y} + iB_{0 \to a}^- e^{-\kappa_M y} - iC_{0 \to a}^+ e^{\tilde{\kappa}_M y} - iC_{0 \to a}^- e^{-\tilde{\kappa}_M y} & 0 < y < a \\ iB_{a \to L}^+ e^{\kappa_G y} + iB_{a \to L}^- e^{-\kappa_G y} - iC_{a \to L}^+ e^{\tilde{\kappa}_G y} - iC_{a \to L}^- e^{-\tilde{\kappa}_G y} & a < y < L \\ iB_{+\infty} e^{-\kappa_G y} - iC_{+\infty} e^{-\tilde{\kappa}_G y} & y > L \end{cases} \tag{13}$$

where $k = \frac{1}{\lambda} = \frac{1}{\sqrt{D\tau}}$, and $\kappa = k\sqrt{1 + i\omega\tau}$. $\tilde{\kappa}$ is the complex conjugate of $\kappa$. The different subscripts M and G refer to the MoS$_2$/graphene heterostructure and graphene channel, respectively. $y = 0$ is at the spot where $I^s_{z\,inj}$ is injected into the channel; $y = a$ is the interface between the MoS$_2$/graphene heterostructure and the graphene channel; $y = L$ is the position of the ferromagnetic electrode. By substituting $\mu^s$ into (8), the expression of spin current in different regions can also be calculated.

### 9.3 Non-Local Voltage Detected by the Ferromagnetic Electrode

The spin accumulation in the spin diffusion channel can be detected as an electrical voltage on a ferromagnetic electrode. For a ferromagnetic material, the electrochemical potential of charge and spin can be described as

$$\mu^{Fc} = \frac{\mu^{F\uparrow}+\mu^{F\downarrow}}{2} \tag{14}$$

$$\mu^{Fs} = \frac{\mu^{F\uparrow}-\mu^{F\downarrow}}{2}. \tag{15}$$

By considering the spin up and spin down as two parallel conduction channels

$$\sigma^{Fc} = \sigma^{F\uparrow} + \sigma^{F\downarrow} \tag{16}$$

$$\sigma^{Fs} = \sigma^{F\uparrow} - \sigma^{F\downarrow} \tag{17}$$

$$P^F_\sigma = \frac{\sigma^{F\uparrow}-\sigma^{F\downarrow}}{\sigma^{F\uparrow}+\sigma^{F\downarrow}} = \frac{\sigma^{Fs}}{\sigma^{Fc}}, \tag{18}$$

the charge and spin current flowing through the ferromagnetic material can be expressed as

$$I^{Fc} = -w \cdot d \cdot \sigma^{Fc}(\nabla\mu^{Fc} + P^F_\sigma \nabla\mu^{Fs}) \tag{19}$$

$$I^{Fs} = -w \cdot d \cdot \sigma^{Fc}(P^F_\sigma \nabla\mu^{Fc} + \nabla\mu^{Fs}), \tag{20}$$

where $d$ is the width of the ferromagnetic electrode. With the device geometry in our model, the ferromagnetic electrode acts as a spin detector, and the net charge current $I^{Fc} = 0$. Combining (19) and (20) to cancel out $\mu^{Fc}$, and utilizing $I^{Fc} = 0$, we derive

$$I^{Fs} = -\left(1 - P^{F2}_\sigma\right) \cdot w \cdot d \cdot \sigma^{Fc}\nabla\mu^{Fs}. \tag{21}$$

The spin electrochemical potential can be described as

$$\mu^{Fs} = \mu^{Fs}(z=0)e^{-k^F z}, \tag{22}$$

where $k = 1/\lambda^F$, and $\lambda^F$ is the spin diffusion length of the ferromagnetic material. This leads to

$$I^{Fs} = \left(1 - P^{F2}_\sigma\right)\frac{w \cdot d \cdot \sigma^{Fc}}{\lambda^F}\mu^{Fs}(z=0) = \left(1 - P^{F2}_\sigma\right)\tilde{R}^{F-1}\mu^{Fs}(z=0). \tag{23}$$

$\tilde{R}^F$ is the spin resistance of the ferromagnetic material.

Next we consider the interface between the ferromagnetic electrode and the spin diffusion channel. The interface normally has finite resistance. The interfacial conductivity can be defined as

$$\Sigma^{Ic} = \Sigma^{I\uparrow} + \Sigma^{I\downarrow} \tag{24}$$

$$\Sigma^{Is} = \Sigma^{I\uparrow} - \Sigma^{I\downarrow} \tag{25}$$

$$P_\sigma^I = \frac{\Sigma^{I\uparrow} - \Sigma^{I\downarrow}}{\Sigma^{I\uparrow} + \Sigma^{I\downarrow}} = \frac{\Sigma^{Is}}{\Sigma^{Ic}}. \tag{26}$$

The charge and spin current across the interface can be written as

$$I^{Ic} = -w \cdot d \cdot \Sigma^{Ic}[(\mu^{Fc}(z=0) - \mu^{Nc}) + P_\sigma^I(\mu^{Fs}(z=0) - \mu^s)] \tag{27}$$

$$I^{Is} = -w \cdot d \cdot \Sigma^{Ic}[P_\sigma^I(\mu^{Fc}(z=0) - \mu^{Nc}) + (\mu^{Fs}(z=0) - \mu^s)]. \tag{28}$$

Considering that the net charge current across the interface $I^{Ic} = 0$, combining the above two equations gives

$$I^{Is} = -\left(1 - P_\sigma^{I\,2}\right) R^{I\,-1}(\mu^{Fs}(z=0) - \mu^s). \tag{29}$$

Considering the continuity of spin current $I^{Is} = I^{Fs}$, and combining equation (23) and equation (29), we derive

$$I^{Fs} = -\left(1 - P_\sigma^{I\,2}\right) R^{I\,-1}(\mu^{Fs}(z=0) - \mu^s) = \left(1 - P_\sigma^{F\,2}\right) \tilde{R}^{F\,-1}\mu^{Fs}(z=0) \tag{30}$$

$$I^{Fs} = \left(\frac{\tilde{R}^F}{1 - P_\sigma^{F\,2}} + \frac{R^I}{1 - P_\sigma^{I\,2}}\right)^{-1} \cdot \mu^s \tag{31}$$

$$\mu^{Fs}(z=0) = \frac{\tilde{R}^F\left(1 - P_\sigma^{F\,2}\right)^{-1}}{\tilde{R}^F\left(1 - P_\sigma^{F\,2}\right)^{-1} + R^I\left(1 - P_\sigma^{I\,2}\right)^{-1}} \mu^s. \tag{32}$$

### 9.4 Boundary Conditions and Determination of the Electrochemical Potentials and Spin Currents

The continuity condition requires that both the electrochemical potential and spin current should be continuous in the two-dimensional channel. With an external magnetic field applied in the y direction, the spin polarization will only precess in the x-z plane, and spin polarization in the y direction is always zero. In the x-z plane, the continuity condition leads to

$$\begin{cases} \mu_{x(y<0)}^s(y=0) = \mu_{x(0<y<a)}^s(y=0) \\ \mu_{z(y<0)}^s(y=0) = \mu_{z(0<y<a)}^s(y=0) \\ \mu_{x(0<y<a)}^s(y=a) = \mu_{x(a<y<L)}^s(y=a) \\ \mu_{z(0<y<a)}^s(y=a) = \mu_{z(a<y<L)}^s(y=a) \\ \mu_{x(a<y<L)}^s(y=L) = \mu_{x(y>L)}^s(y=L) \\ \mu_{z(a<y<L)}^s(y=L) = \mu_{z(y>L)}^s(y=L) \end{cases} \tag{33}$$

$$\begin{cases} I^S_{x(y<0)}(y=0) = I^S_{x(0<y<a)}(y=0) \\ I^S_{z(y<0)}(y=0) = I^S_{z(0<y<a)}(y=0) + I^S_{z\,inj} \\ I^S_{x(0<y<a)}(y=a) = I^S_{x(a<y<L)}(y=a) \\ I^S_{z(0<y<a)}(y=a) = I^S_{z(a<y<L)}(y=a) \\ I^S_{x(a<y<L)}(y=L) = I^S_{x(y>L)}(y=L) + I^{Fs}_x \\ I^S_{z(a<y<L)}(y=L) = I^S_{z(y>L)}(y=L) + I^{Fs}_z \end{cases} \qquad (34)$$

Substituting equation (8), (11) and (13) into the above equations produces 12 linear equations with 12 unknown coefficients: $B_{-\infty}, C_{-\infty}, B^+_{0\to a}, B^-_{0\to a}, C^+_{0\to a}, C^-_{0\to a}, B^+_{a\to L}, B^-_{a\to L}, C^+_{a\to L}, C^-_{a\to L}, B_{+\infty}, C_{+\infty}$. These equations can be solved by linear algebra to determine the values of the 12 coefficients. By inserting these coefficient values into equation (11) and (13), we obtain the spatial dependence of the spin-dependent electrochemical potentials in the spin diffusion channel: $\mu^S_x(y)$ and $\mu^S_z(y)$. Figure 6b shows a plot of these electrochemical potentials. Figure 6c shows a plot of related quantities $\mu^S_x(y)/\mu^S_{total}(y)$ and $\theta_{precession}(y)$. Here, $\mu^S_{total}(y) = \sqrt{[\mu^S_x(y)]^2 + [\mu^S_y(y)]^2 + [\mu^S_z(y)]^2}$ with $\mu^S_y(y) = 0$, and the average precession angle is $\theta_{precession}(y) = \arcsin(\mu^S_x(y)/\mu^S_{total}(y))$. We note that these particular curves in Figure 6b and 6c were obtained through a fitting procedure described in section 9.6.

### 9.5 Non-local Voltage Detected by the Ferromagnetic Electrode

The non-local voltage detected by the ferromagnetic electrode can be defined as

$$V_{NL} = \mu^{Fc}_x(z \to \infty) - \mu^c_x(y \to \infty). \qquad (35)$$

Recall that the charge current in the ferromagnetic detector can be written as

$$I^{Fc} = -w \cdot d \cdot \sigma^{Fc}(\nabla \mu^{Fc} + P^F_\sigma \nabla \mu^{Fs}) = 0, \qquad (36)$$

which indicates

$$\mu^{Fc}_x(z \to \infty) - \mu^{Fc}_x(z=0) = -P^F_\sigma[\mu^{Fs}_x(z \to \infty) - \mu^{Fs}_x(z=0)] = P^F_\sigma \cdot \mu^{Fs}_x(z=0). \qquad (37)$$

In the spin diffusion channel, the electrochemical potential is constant. By substituting (37) and $\mu^{Nc}(y=L) = \mu^{Nc}(y=\infty)$ into (35), we can rewrite the non-local voltage as

$$V_{NL} = \mu^{Fc}_x(y=L, z=0) + P^F_\sigma \cdot \mu^{Fs}_x(y=L, z=0) - \mu^c_x(y=L). \qquad (38)$$

Now recall that the charge current at the ferromagnetic electrode/graphene interface can be written as

$$I^{Ic} = -w \cdot d \cdot \Sigma^{Ic}[(\mu^{Fc}(z=0) - \mu^c) + P^I_\sigma(\mu^{Fs}(z=0) - \mu^s)] = 0, \qquad (39)$$

which indicates

$$\mu^{Fc}(y=L, z=0) - \mu^c(y=L) = -P^I_\sigma(\mu^{Fs}(y=L, z=0) - \mu^s(y=L)). \qquad (40)$$

Notice that from equation (9), the non-local voltage can be rewritten as

$$V_{NL} = -P_\sigma^I \left( \mu_x^{Fs}(y = L, z = 0) - \mu_x^s(y = L) \right) + P_\sigma^F \cdot \mu_x^{Fs}(y = L, z = 0). \quad (41)$$

Combining the above equation with (32) leads to

$$V_{NL} = \left[ \frac{P_\sigma^I R^I \left(1 - P_\sigma^{I\,2}\right)^{-1}}{\tilde{R}^F \left(1 - P_\sigma^{F\,2}\right)^{-1} + R^I \left(1 - P_\sigma^{I\,2}\right)^{-1}} + \frac{P_\sigma^F \tilde{R}^F \left(1 - P_\sigma^{F\,2}\right)^{-1}}{\tilde{R}^F \left(1 - P_\sigma^{F\,2}\right)^{-1} + R^I \left(1 - P_\sigma^{I\,2}\right)^{-1}} \right] \cdot \mu_x^s(y = L). \quad (42)$$

By further assuming that $P_\sigma^I = P_\sigma^F = P_{eff}$, the equation can be simplified as

$$V_{NL} = P_{eff} \cdot \mu_x^s(y = L). \quad (43)$$

We note that the quantity $V_{NL,total} = V_{NL,M+} - V_{NL,M-}$ from the main text is equal to $2V_{NL}$ in equation (43).

### 9.6 Fitting the Experimental Data with the Presented Model

There are six unknowns in our one-dimensional model: optically injected spin current $I_{z\,inj}^s$, diffusion coefficient $D_M$ and spin lifetime $\tau_M$ of the MoS$_2$/graphene heterostructure, diffusion coefficient $D_G$ and spin lifetime $\tau_G$ of the graphene channel, and the effective spin polarization of the detector electrode $P_{eff}$. To reduce the parameter space, we first perform an independent fitting of the Hanle curve from electrical spin injection, as described in section 2. The effective spin polarization of the detector electrode $P_{eff}$, diffusion coefficient $D_G$ and spin lifetime $\tau_G$ of graphene can be extracted from fitting the electrical spin injection data. As discussed in section 2, we obtain $P_{eff} = 0.317$, $\tau_G = 308$ ps, and $D_G = 0.0301$ m$^2$/s. This reduces the free parameters in our model to three: $I_{z\,inj}^s, D_M, \tau_M$.

To determine the three free parameters, we fit the experimental data of $V_{NL}$ vs. $B_y$ (Figure 6a) using a least-squares fitting procedure. To determine the sum of squared residual, we generate a $V_{NL}(B_y)$ curve for given values of $I_{z\,inj}^s, D_M, \tau_M$ as follows. The 12 linear equations in (33) and (34) can be written as a 12-by-12 matrix at a particular applied magnetic field ($B_y$) and solved to determine the 12 coefficients. Substituting these coefficients into equation (11) yields $\mu_x^s(y)$, and equation (42) yields $V_{NL}$ for that particular field. Repeating this calculation for a series of $B_y$ values generates the curve $V_{NL}(B_y)$. From this $V_{NL}(B_y)$ and the experimental data $V_{NL}^{exp}(B_y)$, the sum of squared residual is given by $\sum_i \left( V_{NL}^{exp}(B_y) - V_{NL}(B_y) \right)^2$ where $i$ indexes the data point. This residual is utilized for the fitting algorithm to determine the parameters $I_{z\,inj}^s, D_M, \tau_M$ that minimize the residual. The fitting in Figure 6a of the main text results in the values $D_M = 0.0183\ m^2/s$, $\tau_M = 23.9\ ps$, and $I_{z\,inj}^s = 116\ nA$.

One can estimate the efficiency of optical spin injection based on the spin current of 116 nA. If we assume an absorption of ~5% by monolayer MoS$_2$,[7] then the average power of 100 μW is converted to an absorption of ~1.6 ×10$^{13}$ photons/sec ( = (5%) 100 $\mu$W/$E_{ph}$, where $E_{ph}$ = 1.93 eV). This yields an estimate of ~0.05 spins per photon absorbed $\left(= \frac{116\ nA}{(1.602\times 10^{-19}\ C)(1.6\times 10^{13}\ photon/s)}\right)$. The ~5% efficiency of optical spin injection is comparable to the efficiency of electrical spin injection, typically 1-10%, although higher values (~30%) have been observed.[2] In principle, the optical spin injection efficiency could be substantially higher because the optical selection rules of monolayer TMDs permit 100% spin/valley polarization upon absorption of circularly polarized light.[8, 9] Materials such as monolayer WSe$_2$, with longer spin/valley lifetimes, may prove to be useful for improving the optical spin injection efficiency.[10-12]